# Novel Nongenomic Signaling by Glucocorticoid May Involve Changes to Liver Membrane Order in Rainbow Trout

Laura Dindia[1], Josh Murray[1], Erin Faught[1], Tracy L. Davis[2], Zoya Leonenko[1], Mathilakath M. Vijayan[1]*

1 Department of Biology, University of Waterloo, Waterloo, Ontario, Canada, 2 Department of Biological Sciences, University of Idaho, Moscow, Idaho, United States of America

## Abstract

Stress-induced glucocorticoid elevation is a highly conserved response among vertebrates. This facilitates stress adaptation and the mode of action involves activation of the intracellular glucocorticoid receptor leading to the modulation of target gene expression. However, this genomic effect is slow acting and, therefore, a role for glucocorticoid in the rapid response to stress is unclear. Here we show that stress levels of cortisol, the primary glucocorticoid in teleosts, rapidly fluidizes rainbow trout (*Oncorhynchus mykiss*) liver plasma membranes *in vitro*. This involved incorporation of the steroid into the lipid domains, as cortisol coupled to a membrane impermeable peptide moiety, did not affect membrane order. Studies confirmed that cortisol, but not sex steroids, increases liver plasma membrane fluidity. Atomic force microscopy revealed cortisol-mediated changes to membrane surface topography and viscoelasticity confirming changes to membrane order. Treating trout hepatocytes with stress levels of cortisol led to the modulation of cell signaling pathways, including the phosphorylation status of putative PKA, PKC and AKT substrate proteins within 10 minutes. The phosphorylation by protein kinases in the presence of cortisol was consistent with that seen with benzyl alcohol, a known membrane fluidizer. Our results suggest that biophysical changes to plasma membrane properties, triggered by stressor-induced glucocorticoid elevation, act as a nonspecific stress response and may rapidly modulate acute stress-signaling pathways.

**Funding:** This study was supported by the Natural Sciences and Engineering Research Council (NSERC) of Canada discovery grant and Discovery Accelerator Supplement to MMV. LD was the recipient of the NSERC post-graduate scholarship. AFM infrastructure in Leonenko's lab was supported by CFI and ORF. The funders had no role in study design, data collection and analysis, decision to publish, or preparation of the manuscript.

**Competing Interests:** Dr. Zoya Leonenko is a PLOS ONE Editorial Board member and she is a coauthor on this MS. This does not alter the authors' adherence to all the PLOS ONE policies on sharing data and materials.

* E-mail: matt.vijayan@ucalgary.ca

## Introduction

The neuroendocrine response to stress is highly conserved among vertebrates and essential to reestablish homeostasis [1]. The principal stress hormones, epinephrine and glucocorticoid, have critical functions in the stress adaptation process [2]. The fight-or-flight response involves the activation of the sympathetic nervous system leading to the rapid release of epinephrine from chromaffin cells [2]. Glucocorticoid hormone release occurs in response to activation of the hypothalamus-pituitary-adrenal (HPA) axis [2], and reaches peak levels only after the catecholamine response [1]. Most glucocorticoid effects are mediated by the glucocorticoid receptor (GR), a ligand-bound transcription factor, which regulates protein synthesis [2]. In addition to the slower genomic actions, glucocorticoid also elicits rapid effects that are independent of gene transcription and are broadly categorized as nongenomic signaling [3], but a role for this in stress adaptation is unclear [4].

Nongenomic steroid signaling involves activation of membrane-bound receptors, but a glucocorticoid-specific membrane receptor is yet to be identified [3–5]. In addition to receptor-mediated effects, biophysical changes in lipid bilayers due to steroid intercalation may activate signaling pathways [6,7]. Despite studies showing that changes in plasma membrane properties can rapidly affect the cellular stress response [8], very little is known about the effect of stress steroids on membrane properties and the associated signaling pathways [9]. We tested the hypothesis that acute stress levels of cortisol, the principal glucocorticoid in teleost fishes, rapidly modulate cell stress signaling pathways by altering the biophysical state of the plasma membrane. This was tested using rainbow trout (*Oncorhynchus mykiss*) liver, an ideal model as lipid dynamics and plasma membrane properties have been well characterized [10]. We utilized steady-state fluorescence polarization and atomic force microscopy (AFM) to investigate the effect of cortisol on trout liver plasma membrane properties. Rapid changes in phosphorylation status of putative protein kinase A (PKA), protein kinase C (PKC) and Akt substrate proteins in trout hepatocytes were used to confirm modulation of cell signaling pathways in response to cortisol treatment. Benzyl alcohol, a known membrane fluidizer and initiator of the cellular stress response, was also used as a positive control to assess whether changes in plasma membrane fluidity by cortisol may be involved in the cell signaling events. Our results demonstrate for the first time that stressed levels of cortisol rapidly activate stress-related signaling pathways in rainbow trout hepatocytes. We propose alteration in membrane fluidity as a novel nonspecific glucocor-ticoid-mediated stress response leading to the rapid modulation of stress signaling pathways.

## Materials and Methods

### Animals & Sampling

Juvenile rainbow trout (*Oncorhynchus mykiss*; 100–300 g) purchased from Alma Aquaculture Research Station (Alma, ON) were maintained at the University of Waterloo aquatic facility exactly as described before [11]. The tanks were supplied with a constant flow of aerated well water ($12\pm2°C$) and were maintained under a 12 hL:12 hD photoperiod. Trout were acclimated for at least two weeks prior to experiments and were fed commercial trout feed (Martin Mills, Elmira, ON) to satiety once daily, 5 days a week. Experiments were approved by the University of Waterloo Animal Care Protocol Review Committee and adhere to guidelines established by the Canadian Council on Animal Care for the use of animals in teaching and research.

### Liver Plasma Membrane

Liver plasma membranes were isolated using sucrose gradient as described previously [12]. The membrane pellet was resuspended in TCD buffer (300 mM sucrose, 10 mM Tris-HCl, 1 mM dithiothreitol (DDT), 0.5 mM $CaCl_2$, 1X protease inhibitor cocktail, pH 7.5; Sigma) and frozen at $-70°C$. All steps, including centrifugation, were carried out at $4°C$. The enrichment of the membrane fraction was determined as described previously by measuring the activities of $Na^+/K^+$-ATPase [14] and lactate dehydrogenase [15]. The six-fold higher $Na^+/K^+$-ATPase (H: $1.2\pm0.1$ vs. M: $6.9\pm1.1$; n = 7–8) and thirteen-fold higher 5′- nucleotidase (H: $16\pm1.5$ vs. M: $178\pm48$; n = 7–8) activities (U/g protein) in the membrane (M) fraction compared to the initial tissue homogenate (H), confirm membrane enrichment. The ~90% drop in LDH activity (H: $1055\pm40$ vs. M: $124\pm14$; n = 7–8) in the membrane fraction further confirms enriched plasma membranes with negligible cytosolic contamination.

### Membrane Cortisol and Cholesterol Analysis

Membrane cortisol (after diethyl ether extraction) concentrations were determined by radioimmunoassay exactly as described previously [16]. Diethyl ether extracted plasma membrane cholesterol levels were measured using a cholesterol oxidase enzymatic kit (Wako Chemicals, Richmond, VA).

### DPH Anisotropy

The hepatic plasma membrane fluidity was analyzed by measuring the steady-state fluorescence polarization using the membrane fluorescent probe, 1,6-diphenyl-1,3,5 hexatriene (DPH; Life Technologies Inc., Burlington, ON) exactly as described previously [17]. The degree of fluorescence polarization (anisotropy) is directly related to the mobility of DPH within the lipid environment, therefore, DPH anisotropy is inversely related to membrane fluidity. Membrane samples were added to 96-well opaque plates (Corning Incorporated, New York, USA; approximately 300 ng protein/µl; 100 µl total) and incubated with DPH (1:100 of 4.7 mM stock) for 30 min in the dark. To assess membrane order, isolated plasma membranes (0.5 mg/ml protein) were incubated with control, cortisol (10–1000 ng/ml, n = 6), peptide moiety (PEP; 2.72 µM, n = 5), cortisol-PEP (2.75 µM, n = 5), 17β-estradiol (10 µM, n = 6), testosterone (10 µM, n = 6) or benzyl alcohol (25 mM, n = 7) for 30 min while shaking at room temperature. This peptide conjugate has previously been used to differentiate between intracellular (genomic) versus membrane-mediated (nongenomic) steroid effect [18]. Readings were taken at various temperatures starting at $2°C$ thirty min post-treatment, followed by 12, 24 and $37\pm1°C$. The required temperature (reached within approximately 5–10 min) was maintained and this was confirmed by temperature monitoring within each well, using a digital thermometer, immediately prior to and after the anisotropy measurements. Experiments were conducted on a minimum of five independent trout plasma membrane samples. Mifepristone (a GR antagonist) had its own effect on membrane order (see Figure S1) and, therefore, was not used as a tool for blocking GR effects in the present study.

### Preparation of Cortisol-peptide (Cortisol-PEP) Conjugate

Conjugation of cortisol to form a derivative was carried out as reported by Erlanger et al. [19]. Cortisol-carboxy methyl oxime (Cortisol-CMO (4-pregnen-11b,17,21-triol-3,20-dione3-O-carboxymethyloxime, catalog number Q3888-000) was purchased from Steraloids Inc. (Newport, RI). The peptide conjugated to the CMO is a 15 amino acid sequence of the steroidogenic acute regulatory protein (N-terminus-SGGEVVVDQPMERLY-C-terminus; Proteomics Core Facility, Washington State University, Pullman, WA). The PEP is conjugated via the serine to the CMO using a mixed anhydride technique [19] using *N,N*-dimethylformamide (DMF) as solvent, tri-*N*-butylamine, and isobutyl chloroformate. This conjugation procedure produces a product of 1:1 stoichiometry of a cortisol molecule to a single PEP sequence. The reaction is added to LH-20 Sephadex column to separate the cortisol-PEP, free cortisol, and free PEP. Based on the absorbance at 280 nm, three peaks are derived from the separation on the column with the first peak as cortisol-PEP. This method of obtaining just the hormone conjugate has been confirmed for E2-PEP [20] using Waters QTOF-micro electrospray mass 89 spectometer with the sample introduced by direct infusion (Macromolecular Resources, Colorado State University, Fort Collins, CO).

### Atomic Force Microscopy (AFM)

Plasma membrane surface topography (height changes) and phase (viscoelastic changes) were measured simultaneously using atomic force microscopy (AFM). AFM measurements were carried out in a fluid cell (Molecular Imaging) using the Agilent Technologies 5500 Scanning Probe Microscope in intermittent contact mode (MAC mode) at 0.7 ln/s as described before [21]. Precise force regulation was obtained in MAC mode by using a magnetically coated cantilever (MacLevers Type II from Agilent Technologies; force constant: 2.8 N/m, tip radius: 7 nm, and height: 10–15 µm) Membrane samples were transferred onto a freshly cleaved piece of mica placed within the liquid cell and equilibrated for 10 minutes followed by a quick rinse with nanopure water. Plasma membranes were scanned immediately after rinsing (control, 0 min) or 30 min after incubation with or without cortisol (100 ng/mL). The scan immediately after the rinse was used to ensure that there were no time-dependent effects on plasma membrane topography and phase. Control membranes scanned 30 min after the rinse were compared to cortisol-treated membranes to determine cortisol treatment effects. Scanning at 0.7 ln/s takes approximately 15 min, therefore, the initial scan (referred to as 0 min) was completed within 15 min and the post-treatment scans (referred to as 30 min) within 45 min.

Imaging was repeated with membranes from four independent fish. Images were analyzed using Agilent Image Processing software. Quantitative analysis of surface coverage of higher domains was carried as described previously [22] using ImageJ software (Rasband, W.S., ImageJ, U. S. National Institutes of Health, Bethesda, Maryland, USA; http://rsbweb.nih.gov/ij/).

Briefly, surface coverage was calculated by taking the sum of pixels associated with regions above a certain height (to select for higher domains) divided by the sum of the total number of pixels in the image. Membrane roughness for topography and differences in viscoelasticity was calculated by taking the average difference in height or phase between lower and higher membrane regions, respectively.

## Hepatocyte Experiment

Rainbow trout hepatocytes were isolated using *in situ* collagenase perfusion and maintained exactly as described previously [23]. Hepatocyte viability was >95% and the cells were suspended in L-15 (Sigma, St. Louis, MO) medium and plated in six-well tissue culture plates (Sarstedt, Inc., Newton, NC) at a density of 1.5 million cells/well (0.75 million cells/ml). Cells were maintained at 13°C for 24 h at which time the L-15 media was replaced and the cells were allowed to recover for an additional 2 h before the start of experiments. Cells were treated for 10 min either in the absence (0.01% ethanol) or presence of cortisol (100 or 1000 ng/mL) or benzyl alcohol (25 mM). We were unable to use PEP as a tool to separate membrane receptor-mediated effects of cortisol from changes due to membrane order because PEP by itself affected acute signaling pathways in trout hepatocytes (data not shown). The reaction was stopped by replacing L-15 media with 100 µl ice cold lysis buffer (50 mM Tris, 0.25 M sucrose, 1% SDS, 10 mM NaF, 5 mM EDTA, 5 mM NEM, 0.1% Nonidet-P40). Lysed cells were quickly heated at 95°C for 5 min followed by brief sonication (sonic dismembrator, Fisher Scientific). Experiments were repeated with hepatocytes isolated from three independent fish.

## Immunoblotting

Protein concentration was measured using the bicinchoninic acid (BCA) method using bovine serum albumin as the standard. All samples were diluted in Laemmli's sample buffer (1 M tris-HCl, pH 6.8, 60 mM, glycerol 25%, SDS 2%, β-mercaptoethanol 14.4 mM, bromophenol blue 0.1%). Total protein (40 µg) was separated on a 10% SDS-PAGE and transferred to nitrocellulose membrane and blocked with 5% solution of non-fat dry milk in 1 X TTBS (2 mM Tris, 30 mM NaCl, 0.01% Tween, pH 7.5) for 1 h at room temperature. This was followed with an overnight incubation (1:1000 dilution) with either phospho-(Ser) PKC substrate, phospho-Akt substrate (RXXS/T) or phospho-PKA Substrate (RRXS/T) polyclonal rabbit antibodies (Cell Signaling Technology, Beverly, MA). Blots were incubated for 1 h at room temperature with anti-rabbit horseradish peroxidase (HRP)-labeled secondary antibody (Bio-rad; 1:3000 dilutions in 5% skim milk). Protein bands were detected with ECL Plus™ chemiluminescence (GE Health Care, Baie d'Urfe, QC) and imaged using either the Typhoon 9400 (Amersham Biosciences) or the Pharos FX Molecular Imager (Bio-rad). Total lane or protein band intensity was quantified using AlphaImager HP™ (Alpha Innotech, CA). Equal loading was confirmed by incubation of membranes with Cy3™ conjugated monoclonal mouse β-actin antibody (Sigma, 1:1000) for 1 h at room temperature.

## Statistical Analysis

A one-way or two-way analysis of variance was used for multiple comparisons and a least significant differences (LSD) *post hoc* test was used to compare within factor effects. Statistics were performed either on raw or log transformed data (when necessary to meet normality and equal variance assumptions). A probability level of p<0.05 was considered significant. All statistical analyses were performed using SigmaPlot 11 software (Systat Software Inc., San Jose, CA, USA).

# Results

## Membrane Properties – Cortisol Exposure in vitro

To determine whether cortisol accumulates within plasma membranes, enriched liver plasma membranes from unstressed trout were exposed to stress levels of cortisol (100 ng/mL) for 30 min *in vitro*. Cortisol treatment caused a significant five-fold increase in membrane cortisol content (0.50±0.26 ng/mg protein) compared to untreated membranes (0.11±0.06 ng/mg protein, n = 3, Paired Student's t-test), while membrane cholesterol levels were not significantly different between the two groups (60.9±10.9 µg/mg protein in control versus 55.2±9.4 µg/mg protein in cortisol treated membranes).

**Steady-state fluorescence polarization.** As expected, DPH anisotropy decreased with increasing temperatures (Fig. 1). Benzyl alcohol significantly increased hepatic plasma membrane fluidity compared to the control membrane (Fig. 1A). Exposure to stressed levels of cortisol (100–1000 ng/mL) significantly increased hepatic plasma membrane fluidity, whereas resting level of cortisol (10 ng/mL) reported in trout had no significant effect on fluidity compared to the control group (Fig. 1B). When cortisol was coupled to a peptide moiety (PEP) to make it membrane impermeable (cortisol-PEP), there was no significant effect on membrane fluidity (Fig. 1C). Also, neither pharmacological levels of 17β-estradiol (10 µM) nor testosterone (10 µM) significantly affected trout liver plasma membrane order (Fig. 1D).

**Atomic force microscopy (AFM).** The surface topography of control (Figs. 2A, a,c) membranes and their corresponding cross-section plots (Figs. 2A, b,d) reveal membrane domains within the plasma membrane that differ in height. The solid arrow points to a lower membrane domain (darker regions), while the dotted arrow denotes a higher domain (lighter regions, Fig. 2A, c). The difference in height between the low and high domains (average membrane roughness) of control plasma membranes did not vary over the 30 min incubation (0 min: 2.60 nm±0.073 nm versus 30 min: 2.49 nm±0.11 nm). However, surface topography differed considerably after cortisol treatment (Figs. 2A, e,g, cross-sections Figs. 2A, f,h) compared to control membranes at 30 min (Figs. 2A, a,c, cross-sections Figs. 2A, b,d). In particular, by comparing the cross-sections, maximum roughness was higher for membranes treated with cortisol (3.98 nm±0.13) compared to control membranes (2.49 nm±0.11 nm).

In addition to domain height, the phase image (Fig. 2B), which maps the degree of surface adhesion of the cantilever as it interacts with the surface [24], also indicates that the different domains differ in their relative hardness (viscoelastic properties). As with topography, the control phase images did not change over the 30 min period. Unlike topography, cortisol treatment decreased the degree to which the phase differed between the higher and lower regions (Figs. 2B, e,g) compared to control membranes (Figs. 2B, a,c). Specifically, in control membranes there was a nine-fold difference in the phase image (Fig. 2B, b) between the soft versus the most rigid points, whereas there was only a twofold difference after cortisol treatment (calculated from corresponding cross sections; Fig. 2B, d). As seen in the cross-sectional plots of control (Figs. 2B, b,d) and cortisol (Figs. 2B, f,h) treated membranes, this is due to an increase in the surface adhesion of the lower (fluid) domain, whereas the surface adhesion of the upper domain remained unchanged following cortisol treatment (i.e. phase of lower domains increases, whereas phase of upper domains is unchanged in response to cortisol treatment; Fig. 2C).

Lastly, as seen in both the topography and phase images following cortisol treatment (Figs. 2A and 2B [e,f,g,h]), the micro-domains (lipid rafts) increased in frequency and size and, therefore, cover a greater surface area compared to control (Figs. 2A and 2B [a,b,c,d]) membranes. Specifically, surface coverage of higher domains for control membranes was 51%, whereas the surface coverage of higher domains increased to 66% following cortisol treatment.

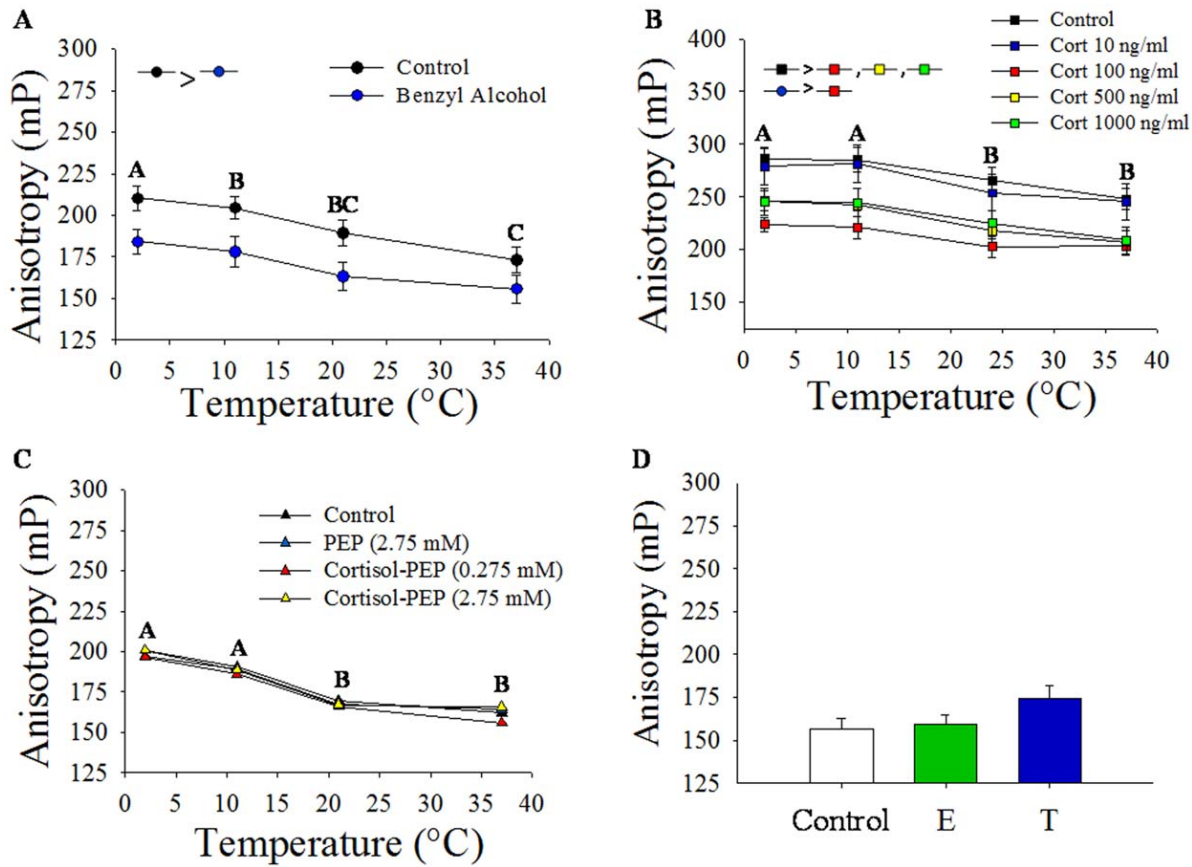

**Figure 1. Cortisol effect on plasma membrane order *in vitro.*** A) 1,6-Diphenyl-1,3,5-hexatriene (DPH) fluorescence anisotropy of enriched hepatic plasma membranes isolated from rainbow trout treated with or without benzyl alcohol for 30 min prior to anisotropy measurement at various temperatures. Data represents mean ± S.E.M (n = 7 independent fish). Different upper case letters indicate significant temperature effects and inset indicates significant treatment effects (two-way repeated measures ANOVA, p<0.05). B) DPH fluorescence anisotropy of enriched hepatic plasma membrane fractions treated with cortisol (0, 10, 100, 500, and 1000 ng/ml) for 30 min at various temperatures. Data represents mean ± S.E.M (n = 6 independent fish). Different upper case letters indicate significant temperature effects and inset indicates significant treatment effects (two-way repeated measures ANOVA, p<0.05). C) DPH fluorescence anisotropy of isolated hepatic plasma membrane fractions treated with the peptide conjugate (PEP, equivalent to 1000 ng/ml, 2.75 µM), or cortisol-conjugated peptide (cortisol-PEP, 0.275 µM and 2.75 µM) for 30 min prior to anisotropy measurement at various temperatures. Data represents mean ± S.E.M (n = 5 independent fish). Different upper case letters indicate significant temperature effects (one-way repeated measures ANOVA, p<0.05). D) DPH fluorescence anisotropy of enriched trout hepatic plasma membrane fractions treated with 17β-estradiol (E; 1 µM) or testosterone (T; 1 µM) for 30 min. Data shown as mean ± S.E.M (n = 6 independent fish).
doi:10.1371/journal.pone.0046859.g001

## Hepatocyte Response

To investigate whether physiochemical changes in the plasma membrane are capable of stimulating intracellular signaling in trout hepatocytes we examined the phosphorylation of PKA, PKC and AKT-substrate proteins in response to benzyl alcohol, a known membrane fluidizer, and cortisol. Our results demonstrate that benzyl alcohol (25 mM) significantly increases phosphorylation of PKC (Fig. 3A), PKA (Fig. 3B) and AKT (Fig. 3C) substrate proteins within 10 min. Similar to the response seen with benzyl alcohol, cortisol at 100 and 1000 ng/mL also significantly increased phosphorylation of putative substrate proteins for all three kinases at 10 min post-treatment (Figs. 3A–C).

## Discussion

This study provides novel insight into the role of glucocorticoid in the acute stress response by demonstrating for the first time that stressed levels of this steroid rapidly alter liver plasma membrane fluidity and modulates cell signaling in a piscine model. Although physical changes to plasma membrane structure are known to be an important initiator of intracellular signaling events in response to stressors [8], a role for steroids in this process is far from clear. Our results highlight a hitherto unknown role for cortisol in acute stress adaptation that is nonspecific and involves changes in membrane fluidity.

We demonstrate a rapid change in liver plasma membrane fluidity in response to stressed levels of cortisol *in vitro*. The fluidity changes seen with cortisol were not dose-related, but occurred above a certain threshold suggesting a receptor-independent mechanism likely associated with steroid incorporation into the lipid domain. This was supported by the inability of membrane impermeable cortisol-PEP to alter membrane fluidity. While changes in plasma membrane cholesterol levels alter lipid order

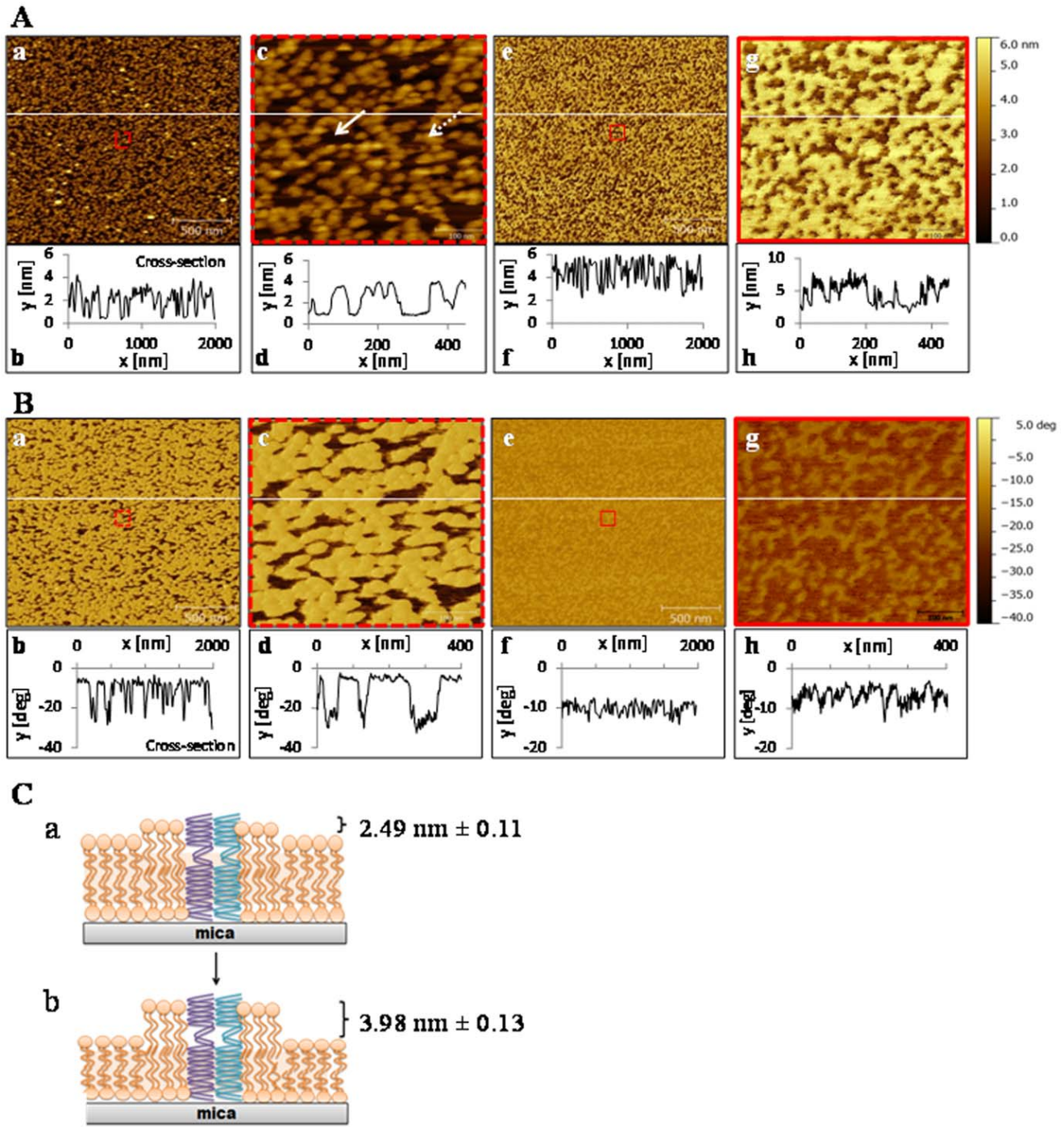

**Figure 2. Cortisol effect on liver plasma membrane topography and surface adhesion.** A) Representative atomic force microscopy (AFM) images of supported hepatic plasma membrane topography (membrane height). Images were taken prior to (a) and 30 min after cortisol (100 ng/mL) treatment (e) in liquid cell at room temperature. A zoomed in scan is also shown for the control (c) and cortisol treated (g) membranes that was scanned for 60 min. The approximate scan region of the zoomed in image is indicated by the dashed red box in the control image (a) and solid red box in the cortisol-treated image (e). Two distinct domains, which differ in height, are visible in both control and cortisol-treated membranes. A representative higher domain is indicated by the dotted arrow, while the lower domain is indicated by the solid arrow (c). Short-term cortisol treatment altered the topography of the plasma membrane. The cross-section graph featured below each image was calculated from points along the white horizontal line. The y-axis represents vertical height (nm), whereas the x-axis represents the horizontal distance (nm). B) Representative AFM images of supported hepatic plasma membrane phase (surface adhesion properties). Images were taken prior to (a) and 30 min after cortisol (100 ng/mL) treatment (e) in liquid cell at room temperature. A zoomed in scan is also shown of the control (c) and cortisol treated (g) membranes that was scanned for 60 min. The approximate scan region of the zoomed in image is indicated by the dashed red box in the control image (a) and solid red box in the cortisol-treated image (e). Two distinct domains, which differ in their viscoelastic (surface adhesion) are visible in both control and cortisol-treated membranes. Acute cortisol treatment altered the viscoelastic properties of the plasma membrane within 30 min of treatment. The cross-section graph featured below each image was calculated from points along the white horizontal line. The y-axis represents degree of deflection (degrees), whereas the x-axis represents the horizontal distance (nm). C) A schematic representation of cortisol's effect on plasma membrane

properties. Short-term incubation with cortisol (b) increased surface roughness (height difference between higher and lower domains) compared to control membrane (a).
doi:10.1371/journal.pone.0046859.g002

[25] that appears unlikely in the present case as membrane cholesterol remained unchanged in response to cortisol treatment. The cortisol-induced fluidization of liver plasma membrane

appears to be steroid specific, as neither 17β-estradiol nor testosterone treatment showed a similar response in trout plasma membrane. This agrees with the recent findings that the chemical structure of the steroid backbone affect interaction with the lipid bilayer and subsequent changes in plasma membrane fluidity [6]. It remains to be determined whether the membrane biophysical effect is also seen with other corticosteroids and not just cortisol. However, cortisol is the primary corticosteroid that is released into the circulation in response to stress in trout. The membrane fluidizing effect of cortisol seen in liver may be a generalized response affecting all tissues in response to stress. Mammalian studies reported a fluidizing effect of glucocorticoid on fetal rat liver [26] and dog synaptosomal membranes [27], whereas an ordering effect was observed in rat renal brush border [28] and rabbit cardiac muscle [29]. This suggests that stress-mediated cortisol effect on membrane order may be tissue-specific, but this remains to be determined in fish. Altogether, our results indicate that stress-induced elevation in cortisol levels rapidly fluidizes liver plasma membrane in rainbow trout.

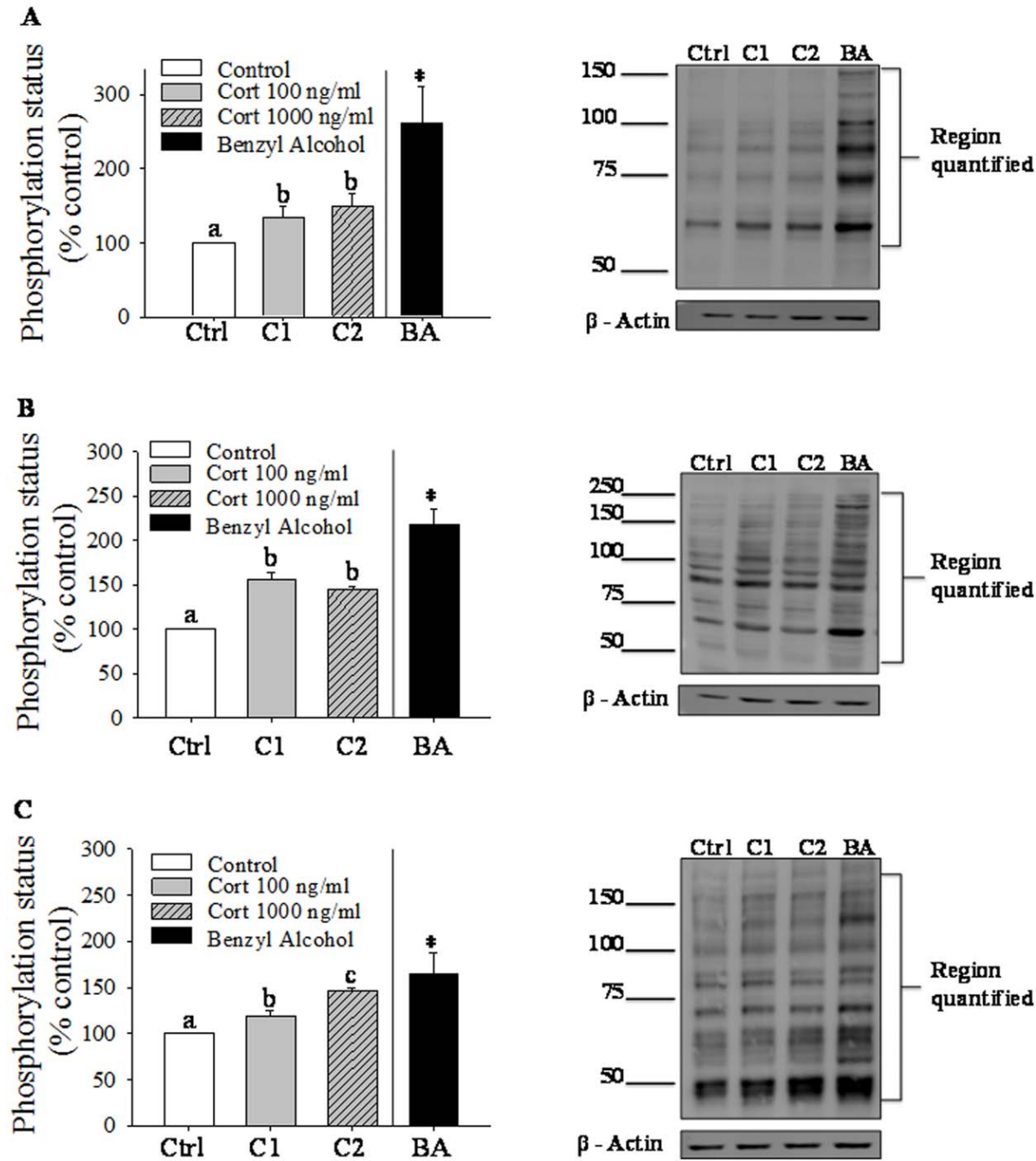

Figure 3. Cortisol effect on rapid cell signaling in trout hepatocytes. Rainbow trout hepatocytes were incubated either with cortisol (0, 100 or 1000 ng/mL) or benzyl alcohol (BA; 25 mM) for 10 min. Cell homogenates (40 μg protein) were probed with polyclonal rabbit antibody (Cell Signaling Technology, Beverly, MA) to either phospho-(Ser) PKC substrate (A), phospho-PKA Substrate (RRXS/T) (B) or phospho-Akt substrate (RXXS/T) (C). Equal loading was confirmed with β-actin (monoclonal mouse antibody; Sigma, St. Louis, MO). A representative immunoblot for each is shown; values are plotted as % control and shown as mean ± S.E.M (n = 3 independent fish); bars with different letters are significantly different (repeated measures ANOVA, p<0.05). *significantly different from control (Paired Student's t-test; p<0.05).
doi:10.1371/journal.pone.0046859.g003

AFM topographical and phase images further indicate that cortisol alters biophysical properties of liver plasma membranes. Specifically, cortisol exposure led to the reorganization of discrete microdomains, likely gel phase (higher domains) and disordered fluid-phase (lower domains) in the lipid bilayer. These discrete domains differed in height, which increased after cortisol treatment. A recent study on erythrocytes also reported a glucocorticoid-induced domain reorganization, which involved formation of large protein-lipid domains by hydrophobic and electrostatic interactions leading to alteration in membrane structure and elasticity [30]. Similar domain changes have also been reported for synthetic lipids in response to halothane exposures or melting transitions [31], treatments that are known to increase membrane fluidity [31,32]. Cortisol appears to have a greater effect on lower domains, as indicated by the greater change in surface adhesion (phase) following steroid treatment, compared to the higher lipid domains. Collectively, stressed levels of cortisol rapidly alter the biophysical properties of trout hepatic plasma membrane. We hypothesize that changes in membrane order by cortisol is the result of a non-uniform fluidization at the nanoscale among different membrane domains.

Rapid changes to membrane order by cortisol may play a role in triggering acute stress-related signaling pathways. Indeed membrane order perturbations lead to rapid activation of cell signaling pathways, including protein kinases [8]. In agreement, benzyl alcohol, a known membrane fluidizer, rapidly induced phosphorylation of PKA, PKC and AKT putative substrate proteins. The intracellular effect of benzyl alcohol has been attributed to its direct effect on plasma membrane structure. Interestingly, cortisol exposure also induced phosphorylation of PKA, PKC and AKT putative substrate proteins as seen with benzyl alcohol, supporting a rapid stress signaling event mediated by changes to membrane order. While membrane receptor mediated nongenomic glucocorticoid signaling has been reported before [9], to our knowledge this is the first report of membrane biophysical changes initiating rapid signaling event induced by stressed levels of cortisol in any animal model.

To date, the genomic effects of cortisol have been the primary focus in establishing the role of this steroid in the acute stress response [33,34]. In liver, stress-induced cortisol has been shown to modulate expression of genes involved in intermediary metabolism, including gluconeogenesis,that is essential for mobilizing glucose to cope with the enhanced energy demand [33–36]. This genomic response to cortisol is slow acting and, therefore, not considered to be important in the rapid glucose regulation associated with the fight-or-flight response [37]. The PKA and AKT [38] signaling pathways are both known to regulate hepatic glucose metabolism, while PKC has been implicated in hepatic insulin resistance [39]. Consequently, cortisol-mediated changes in membrane fluidity may be a key nonspecific stress response triggering the phosphorylation of putative protein kinase substrate proteins. This rapid activation of stress-related signaling pathways by cortisol may be playing an important role in the metabolic adjustments to the fight-or-flight response. As plasma membrane order can affect membrane receptor function [40], we hypothesize that cortisol-induced biophysical membrane changes may also modify hepatocyte responsiveness to other stress signals, including glucoregulatory hormone stimulation. In support, studies have shown a permissive effect of cortisol treatment on epinephrine-mediated glucose production in trout hepatocytes [35,41].

Altogether, our results underscore a novel plasma membrane response to stressed levels of glucocorticoid exposure, leading to a nongenomic signaling event in trout hepatocytes. This rapid and nonspecific cortisol effect may act either alone and/or in concert with membrane receptor activation, to modulate stress-related signaling pathways. We propose that the rapid cortisol-mediated changes in membrane fluidity occur in a non-uniform domain-like manner and may have important consequences to non-specific cellular stress response and adaptation to subsequent stressor insult in animals.

## Supporting Information

**Figure S1 Effect of cortisol, RU486, benzyl alcohol & DMSO on membrane fluidity.** Anisotropy of isolated hepatic membranes with cortisol (1 µM) RU486 (1 µM) combination treatment (RU+CORT; both 1 µM) benzyl alcohol (BOH; 5 mM), dimethyl sulphoxide (DMSO, 2% v/v) or without (control) at both 4°C and 23°C. Values are shown as % control and bars represent means ± S.E.M. (N = 3–9 independent membrane preparations).
(DOCX)

## Author Contributions

Conceived and designed the experiments: LD JM MMV. Performed the experiments: LD JM. Analyzed the data: LD JM EF ZL. Contributed reagents/materials/analysis tools: TLD ZL MMV. Wrote the paper: LD JM EF TLD ZL MMV.

## References

1. Sapolsky RM, Romero LM, Munck AU (2000) How do glucocorticoids influence stress responses? integrating permissive, suppressive, stimulatory, and preparative actions. Endocr Rev 21: 55–89.

2. Charmandari E, Tsigos C, Chrousos G (2005) Endocrinology of the stress response 1. Annu Rev Physiol 67: 259–284.

3. Borski RJ, Hyde GN, Fruchtman S (2002) Signal transduction mechanisms mediating rapid, nongenomic effects of cortisol on prolactin release. Steroids 67: 539–548.

4. Tasker JG, Herman JP. (2011) Mechanisms of rapid glucocorticoid feedback inhibition of the hypothalamic-pituitary-adrenal axis. Stress 14: 398–406.

5. Groeneweg FL, Karst H, de Kloet ER, Joels M (2011) Rapid non-genomic effects of corticosteroids and their role in the central stress response. J Endocrinol 209: 153–167.

6. Rog T, Stimson LM, Pasenkiewicz-Gierula M, Vattulainen I, Karttunen M (2008) Replacing the cholesterol hydroxyl group with the ketone group facilitates sterol flip-flop and promotes membrane fluidity. J Phys Chem B 112: 1946–1952.

7. Lösel R, Wehling M (2003) Nongenomic actions of steroid hormones. Nat Rev Mol Cell Biol 4: 46–55.

8. Vigh L, Nakamoto H, Landry J, Gomez-Munoz A, Harwood JL, et al. (2007) Membrane regulation of the stress response from prokaryotic models to mammalian cells. Ann N Y Acad Sci 1113: 40–51.

9. Borski RJ (2000) Nongenomic membrane actions of glucocorticoids in vertebrates. TEM 11: 427–436.

10. Zehmer JK, Hazel JR (2004) Membrane order conservation in raft and non-raft regions of hepatocyte plasma membranes from thermally acclimated rainbow trout. Biochim Biophys Acta 1664: 108–116.

11. Sandhu N, Vijayan MM (2011) Cadmium-mediated disruption of cortisol biosynthesis involves suppression of corticosteroidogenic genes in rainbow trout. Aquat Toxicol 103: 92–100.

12. Sulakhe SJ (1987) Hepatic adrenergic receptors in the genetically diabetic C57 BL/KsJ (db/db) mouse. Int J Biochem 19: 1181–1186.

13. McGuire A, Aluru N, Takemura A, Weil R, Wilson JM, et al. (2010) Hyperosmotic shock adaptation by cortisol involves upregulation of branchial osmotic stress transcription factor 1 gene expression in mozambique tilapia. Gen Comp Endocrinol 165: 321–329.

14. Solyom A, Trams EG (1972) Enzyme markers in characterization of isolated plasma membranes. Enzyme 13: 329–372.

15. Gravel A, Wilson JM, Pedro DFN, Vijayan MM (2009) Non-steroidal anti-inflammatory drugs disturb the osmoregulatory, metabolic and cortisol responses associated with seawater exposure in rainbow trout. Comp Biochem Physiol C 149: 481–490.

16. Alsop D, Ings JS, Vijayan MM, Earley RL (2009) Adrenocorticotropic hormone suppresses gonadotropin-stimulated estradiol release from zebrafish ovarian follicles. PloS One 4: e6463.

17. Katynski A, Vijayan M, Kennedy S, Moon T (2004) 3, 3′, 4, 4′, 5-pentachlorobiphenyl (PCB 126) impacts hepatic lipid peroxidation, membrane fluidity and β-adrenoceptor kinetics in chick embryos. Comp Biochem Physiol C 137: 81–93.

18. Nagler JJ, Davis TL, Modi N, Vijayan MM, Schultz I (2010) Intracellular, not membrane, estrogen receptors control vitellogenin synthesis in the rainbow trout. Gen Comp Endocrinol 167: 326–330.

19. Erlanger BF, Borek F, Beiser SM, Lieberman S (1957) Steroid-protein conjugates. J Biol Chem 228: 713–728.

20. Davis TL, Whitesell JD, Cantlon JD, Clay CM, Nett TM (2011) Does a nonclassical signaling mechanism underlie an increase of estradiol-mediated gonadotropin-releasing hormone receptor binding in ovine pituitary cells? 1. Biol Reprod 85: 770–778.

21. Moores B, Drolle E, Attwood SJ, Simons J, Leonenko Z (2011) Effect of surfaces on amyloid fibril formation. PloS One 6: e25954.

22. Hane F, Drolle E, Leonenko Z (2010) Effect of cholesterol and amyloid-β peptide on structure and function of mixed-lipid films and pulmonary surfactant BLES: An atomic force microscopy study. Nanomedicine 6: 808–814.

23. Sathiyaa R, Campbell T, Vijayan MM (2001) Cortisol modulates HSP90 mRNA expression in primary cultures of trout hepatocytes. Comp Biochem Physiol B 129: 679–685.

24. Magonov S, Elings V, Whangbo MH (1997) Phase imaging and stiffness in tapping-mode atomic force microscopy. Surf Sci 375: L385–L391.

25. Lingwood D, Simons K (2010) Lipid rafts as a membrane-organizing principle. Science 327: 46–50.

26. Kapitulnik J, Weil E, Rabinowitz R (1986) Glucocorticoids increase the fluidity of the fetal-rat liver microsomal membrane in the perinatal period. Biochem J 239: 41.

27. Deliconstantinos G (1985) Cortisol effect on (na K )-stimulated ATPase activity and on bilayer fluidity of dog brain synaptosomal plasma membranes. Neurochem Res 10: 1605–1613.

28. Levi M, Shayman JA, Abe A, Gross SK, McCluer RH, et al. (1995) Dexamethasone modulates rat renal brush border membrane phosphate transporter mRNA and protein abundance and glycosphingolipid composition. J Clin Invest 96: 207.

29. Gerritsen ME, Schwarz SM, Medow MS (1991) Glucocorticoid-mediated alterations in fluidity of rabbit cardiac muscle microvessel endothelial cell membranes: Influences on eicosanoid release. BBA Biomemb1065: 63–68.

30. Panin LE, Mokrushnikov PV, Kunitsyn VG, Zaitsev BN (2010) Interaction mechanism of cortisol and catecholamines with structural components of erythrocyte membranes.J Phys Chem B 114: 9462–9473.

31. Leonenko ZV, Cramb DT (2004) Revisiting lipid general anesthetic interactions (I): Thinned domain formation in supported planar bilayers induced by halothane and ethanol. Can J Chem 82: 1128–1138.

32. Leonenko Z, Finot E, Ma H, Dahms T, Cramb D (2004) Investigation of temperature-induced phase transitions in DOPC and DPPC phospholipid bilayers using temperature-controlled scanning force microscopy. Biophys J 86: 3783–3793.

33. Vegiopoulos A, Herzig S (2007) Glucocorticoids, metabolism and metabolic diseases. Mol Cell Endocrinol 275: 43–61.

34. Aluru N, Vijayan MM (2009) Stress transcriptomics in fish: A role for genomic cortisol signaling. Gen Comp Endocrinol 164: 142–150.

35. Mommsen TP, Vijayan MM, Moon TW (1999) Rev Fish Biol Fish 9(3): 211–268.

36. Rose AJ, Vegiopoulos A, Herzig S (2010) Role of glucocorticoids and the glucocorticoid receptor in metabolism: Insights from genetic manipulations. J Steroid Biochem Mol Biol 122: 10–20.

37. McEwen BS (2007) Physiology and neurobiology of stress and adaptation: Central role of the brain. Physiol Rev 87: 873–904.

38. Klover PJ, Mooney RA (2004) Hepatocytes: Critical for glucose homeostasis. Int J Biochem Cell Biol 36: 753–758.

39. Samuel VT, Liu Z, Wang A, Beddow SA, Geisler JG, et al. (2007) Inhibition of protein kinase cepsilon prevents hepatic insulin resistance in nonalcoholic fatty liver disease. J Clin Invest 117: 739.

40. Chachisvilis M, Zhang YL, Frangos JA (2006) G protein-coupled receptors sense fluid shear stress in endothelial cells. Proc Natl Acad of Sci USA 103: 15463–15468.

41. Reid S, Moon T, Perry S (1992) Rainbow trout hepatocyte beta-adrenoceptors, catecholamine responsiveness, and effects of cortisol. Amer J Physiol 262: R794–R799.